\documentclass[twocolumn,prb,showpacs,multicol,amsmath,amssymb]{revtex4}
\usepackage[dvips]{graphicx}
\usepackage{subfigure}
\usepackage{graphicx}
\usepackage{dcolumn}
\usepackage{bm}
\usepackage{graphics}
\usepackage{epsfig,color}

\newcommand{\be}{\begin{equation}}
\newcommand{\ee}{\end{equation}}
\newcommand{\bea}{\begin{eqnarray}}
\newcommand{\eea}{\end{eqnarray}}

\begin{document}
\title{Investigation of frustrated and dimerized classical Heisenberg chains}
\author{J. Vahedi$^1$, S. Mahdavifar$^2$, M. R. Soltani$^3$, M. Elahi$^1$ }
\affiliation{ $^1$  Department of Physics, Science and Research
Branch, Islamic Azad University, Tehran, Iran\\
$^2$Department of Physics, University of
Guilan,41335-1914, Rasht, Iran\\
$^3$Department of Physics, Shahre-Ray branch, Islamic Azad University, Tehran, Iran.}
\date{\today}

\begin{abstract}
We have considered the 1D dimerized frustrated antiferromagnetic (ferromagnetic) Heisenberg model with arbitrary spin $S$. The exact classical magnetic phase diagram at zero temperature is  determined using the LK cluster method. Cluster  method  results, show that the classical ground state phase  diagram of the model is very rich including first and  second-order phase transitions. In the absence of the  dimerization, a second-order phase transition occurs between  antiferromagnetic (ferromagnetic) and spiral phases at  the critical frustration $\alpha_c=\pm 0.25$. In the vicinity  of the critical points $\alpha_c$, the exact classical  critical exponent of the spiral order parameter is found $1/2$.  In the case of dimerized chain ($\delta\neq0$), the spiral order shows stability and exists in some part of the ground state phase diagram. We have found two first-order critical lines in the ground state phase diagram. These critical lines separate the antiferromagnetic from spiral phase.
\end{abstract}

\maketitle

\section{Introduction}\label{sec1}

During the last decades several classical techniques such as the well-known Luttinger-Tisza method\cite{Luttinger46}, vertex model\cite{Baxter} and so on have been introduced to solve classical Hamiltonian exactly. The Luttinger-Tisza method is more effective in systems with bilinear interactions and vertex model usually applied for treating frustrated model\cite{Lieb,Sutherland}.

In a very recent work\cite{TK2009}, T. Kaplan have used a kind of cluster method, hereafter simply LK method,
which is based on a block of three spins to solve frustrated classical Heisenberg model in one dimension
with added nearest neighbor biquadratic exchange interactions. He asserted that the LK method is not limited to one
dimension or to translationally invariant spin Hamiltonians\cite{Lyons64} and expanded his approach to determine the phase diagram of frustrated classical Heisenberg and XY models with added nearest neighbor biquadratic exchange interactions in $d=2$ dimension\cite{LX2010}. In order to check the validity of Kaplan's
phase diagram conjecture, we have investigated his model\cite{TK2009} form quantum point of view for spin-$\frac{1}{2}$ with an accurate algorithm (Lanczos method), and our results, which will be presented elsewhere\cite{JaSaee}, showed that LK method, albeit  is a classical approach but has the
capability to work for aspects of a quantum treatment.

Actually, these are our stimulating reasons to take a quite well known  frustrated and dimerized Heisenberg model and determine its classical ground state phase diagram exactly with strong, but not well known LK method which is able to solve problems rigorously\cite{TK2009}. Let us start with definition  of the dimerized and
frustrated Heisenberg model as follow
\begin{equation}
\emph{H}=J_{1}\sum_{n}\Big[1+(-1)^n\delta\Big]\textbf{S}_{n}\cdot\textbf{S}_{n+1}+J_{2}\sum_{n}\textbf{S}_{n}\cdot\textbf{S}_{n+2},
\label{e1}
\end{equation}
where $\vec{S}_{n}$ is the $n$th classical vector of the length $S$. A spin system is frustrated when the global order because of the competition of different kinds of interaction is incompatible with the local order, so chain with both antiferromagnetic-antiferromagnetic exchanges $(J_{1}>0, J_{2}>0)$ and ferromagnetic-antiferromagnetic exchanges $(J_{1}<0, J_{2}>0)$, hereafter simply AF-AF and F-AF respectively, are frustrated.

Quantum study of this model is well done for the spin-1/2 and spin-1 chains\cite{Shastry81,Bouzerar98, Nakamura97, Kumar07, Chitra95, Controzzi05, Pati96, Oshikawa}. It is found that the quantum fluctuations play a very important role at zero temperature in the ground state phase diagram of the models. This model shows a dimerization transition at $\delta=\delta_{c}$.

For system with spin half, $S = 1/2$,  $\delta_{c}=0$ is the transition point, which is  related to the Lieb-Schultz-Mattis theorem (states that system should be in either twofold degeneracy or gapless excitations  of the ground states at $\delta=0$). Indeed, for AF-AF case and on the undimerized link $\delta=0$, there is a critical frustration parameter $\alpha_{c}=0.2411$\cite{Haldane82,Okamoto92}.  The dimerization transition at $\delta=\delta_{c}=0$ is of second order and system for $\alpha<\alpha_{c}$ shows a gapless Tomonaga-Luttinger Liquid (TLL).  But, for $\alpha>\alpha_{c}$, the ground state is doubly degenerate, showing a spontaneous dimerization. This is a signature of  a first-order dimerization transition at $\delta=\delta_{c}=0$.
  
 On the other hand, system with spin one, $S = 1$, with $\delta=0$  and small $\alpha=\frac{J_2}{J_1}$  lives in Haldane phase and does not represent a transition line.  In contrast, dimerization transition between the Haldane phase and the dimerized phase happens\cite{Affleck87,Kolezhuk96} at a finite $\delta_{c}$, which depends on the $\alpha$ (know as frustration parameter). As matter of fact, for system with spin one, there is a critical frustration $\alpha_{c}$ point which is second order form of transition. This critical point separates a TLL phase for $\alpha<\alpha_{c}$ from first order one for $\alpha>\alpha_{c}$.

\begin{figure}[t]
\includegraphics[width=0.95\columnwidth]{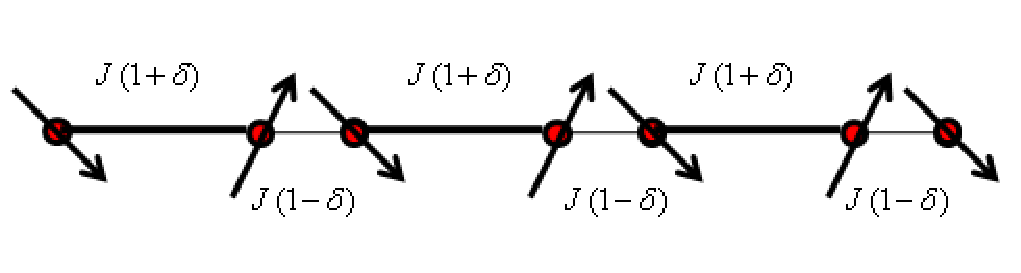}
\caption{(Color online.) The schematic picture of dimerized spins chain.}
\label{dimer}
\end{figure}

But there is not a classical clear picture of different ground state phases of the mentioned model. Having a classical picture, from one hand help us to know that quantum fluctuations destroy which one of the classical orderings. On the other hand for arbitrary large spin model, the classical picture is the same with the quantum picture of the ground state phase diagram. In this work we focus on the 1D frustrated and dimerized systems with arbitrary spin $S$ (see FIG.~\ref{dimer}).  To find the exact classical ground state phase diagram of the model, the LK cluster
method is used. In the absence of the dimerization, by increasing the frustration a classical phase transition occurs at $\alpha_{c}=+0.25~(-0.25)$ from the antiferromagnetic (ferromagnetic) phase into the  spiral magnetic phase.
Our results show that the dimerization parameter induces new magnetic phases including stripe-antiferromagnetic
phase (or \textit{uud} and \textit{duu} phases). Existing of these magnetic phases is independent of length
of spins.

The outline of the paper is as follows. In forthcoming section  we will extensively explain the LK method with
implementing it to our model and in the section III we will summarize our results.
\section{the LK cluster method} \label{sec2}

\begin{figure}[t]
\includegraphics[width=0.95\columnwidth]{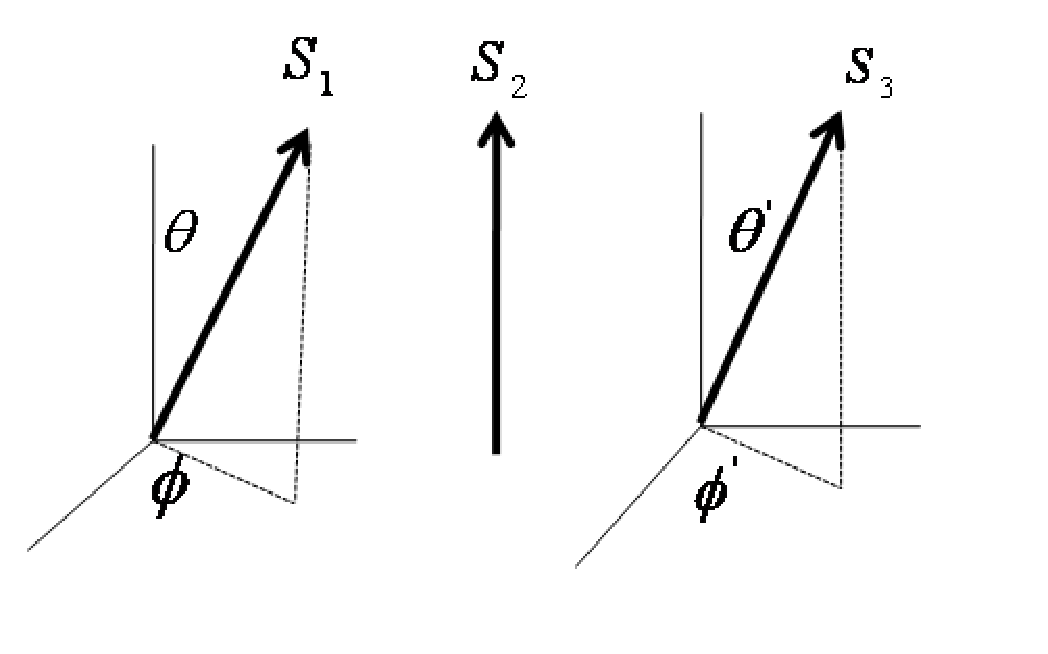}
\caption{Spins cluster configuration.}
\label{angel}
\end{figure}

In order to implement LK method we follow exactly the procedure
in Ref. [2]. Without losing the generality and setting periodic boundary conditions,
Eq.(\ref{e1}) can be rewritten as:
\begin{equation}
\emph{H}_{c}=\sum_{i}h_{c}\big(\textbf{S}_{i-1},\textbf{S}_{i},\textbf{S}_{i+1}\big),
\label{e2}
\end{equation}
where the "cluster energy" involve three neighboring spins is
\begin{eqnarray}
h_{c}(\textbf{S}_{1}, \textbf{S}_{2}, \textbf{S}_{3})&=& \frac{J_{1}}{2}\Big\{\big(1-\delta\big)\textbf{S}_{1}\cdot\textbf{S}_{2}+\big(1+\delta\big)\textbf{S}_{2}\cdot\textbf{S}_{3})\Big\}\nonumber\\
&+&J_{2}(\textbf{S}_{1}\cdot\textbf{S}_{3}). \label{e3}
\end{eqnarray}
It is clear that
\begin{equation}
\emph{H}_{c}\geq\sum_{j}min~h_{c}(\overrightarrow{S}_{j-1},\overrightarrow{S}_{j},\overrightarrow{S}_{j+1}).
\label{cluster-3}
\end{equation}
To minimize $h_{c}$  respect spins directions,
 we first consider coplanar spins,  and label the angles $\theta$, $\theta'$
 made by the end spins with the central spin (see FIG.~\ref{angel}) which
 in coplanar case we set $\phi=0,\phi'=0$. The cluster energy is given by

\begin{eqnarray}
h_{c}(\theta,\theta')&=&S^{2}\Big\{\big(\frac{1-\delta}{2}\big)\cos\theta+\big(\frac{1+\delta}{2}\big)\cos\theta'\nonumber\\
&+&\alpha\cos(\theta-\theta')\Big\}, \label{e4}
\end{eqnarray}
where $\alpha=J_{2}/J_{1}$. Minimizing $h_{c}$
respect $\theta$, $\theta'$ gives the following equation:
\begin{eqnarray}
\frac{\partial h_{c}}{\partial \theta}=-\frac{S^{2}}{2}\Big[(1-\delta)\sin\theta+2\alpha\sin(\theta-\theta')\Big]\nonumber\\
\frac{\partial h_{c}}{\partial
\theta'}=-\frac{S^{2}}{2}\Big[(1+\delta)\sin\theta'-2\alpha\sin(\theta-\theta')\Big].
\label{e5}
\end{eqnarray}
Let's first deal with a case without dimerization, by setting
$\delta=0$ in Eq. (\ref{e5}) we have
\begin{eqnarray}
\frac{\partial h_{c}}{\partial \theta}=-\frac{S^{2}}{2}\Big[\sin\theta+2\alpha\sin(\theta-\theta')\Big]\nonumber\\
\frac{\partial h_{c}}{\partial
\theta'}=-\frac{S^{2}}{2}\Big[\sin\theta'-2\alpha\sin(\theta-\theta')\Big].
\label{e6}
\end{eqnarray}
its solutions are
\begin{eqnarray}
(\theta,\theta')&=&(0,0),(\pi,\pi),(\pi,0),(0,\pi),\nonumber\\
(\theta,\theta')&=&(\theta_{0},-\theta_{0}),~~~\text{(Spiral- type)},~~~ \text{where}\nonumber\\
\cos\theta_{0}&=&-\frac{1}{4\alpha}\rightarrow|\alpha|\geq\frac{1}{4}\nonumber.\\
\label{e7}
\end{eqnarray}

The solutions $(\pi,\pi), (0,0)$ are related to collinear
antiferromagnetic and ferromagnetic states
 respectively. The antiferromagnetic (ferromagnetic) state will minimize the energy in the case of $J_{1}>0$ ($J_{1}<0$).
Solutions $(\pi,0)\rightarrow (\downarrow,\uparrow,\uparrow)$ and
$(0,\pi)\rightarrow (\uparrow,\uparrow,\downarrow)$ are degenerate
states and show spins propagate in the down-up-up and up-up-down respectively \cite{Lyons64}.
Spiral state $(\theta_{0},-\theta_{0})$ with uniform
rotation is also degenerate state. In following we present results of the antiferromagnetic case $J>0$.

By setting minimization conditions
into the Eq.~(\ref{e4}) we have the following energies:
\begin{eqnarray}
h_{\text{antiferro}}&=&h_{c}(\pi, \pi)=S^{2}(-1+\alpha),\nonumber\\
h_{\text{uudd}}&=&h_{c}(0, \pi)=S^{2}(-\alpha),\nonumber\\
h_{\text{spiral}}&=&h_{c}(\theta_{0}, -\theta_{0})=S^{2}(-\frac{1}{8\alpha}-\alpha).
\label{e8}
\end{eqnarray}

By equating these energies in pairs we have found only one
critical point, $\alpha_{c}=0.25$. Because of the continuity in
the derivative $\frac{\partial h_{c}}{\partial \alpha}$, a
second-order phase transition occurs when passing trough
$\alpha=0.25$. The ground state is in the antiferromagnetic phase
in the region of the frustration $\alpha<0.25$ and in the spiral
phase in region  $\alpha>0.25$. In general, the antiferromagnetic
phase is recognized by the non-zero value of the Neel order
parameter defined as
\begin{eqnarray}
M_{st}^{z}=\frac{1}{N} \sum_{n} (-1)^{n}S_{n}^{z}, \label{Neel}
\end{eqnarray}
and the spiral phase in the ground state phase diagram of the spin
systems is characterized by the nonzero value of the spiral order
parameter
\begin{eqnarray}
\chi=\frac{1}{N} \sum_{n} |{\bf S}_{n}\times {\bf S}_{n+1}|.
\label{chirality}
\end{eqnarray}
Using Eq.~(\ref{e7}) we have found the spiral order parameter as
\begin{eqnarray}
\chi&=&0~~~~~~~~~~~~~~~~~~,~~~\alpha<\alpha_{c},\nonumber \\
\chi&=&S^2\sqrt{1-\frac{\alpha_{c}^{2}}{\alpha^{2}}}~~~~,~~~\alpha>\alpha_{c}.
\label{chirality}
\end{eqnarray}

\begin{figure}[t]
\includegraphics[width=1.1\columnwidth]{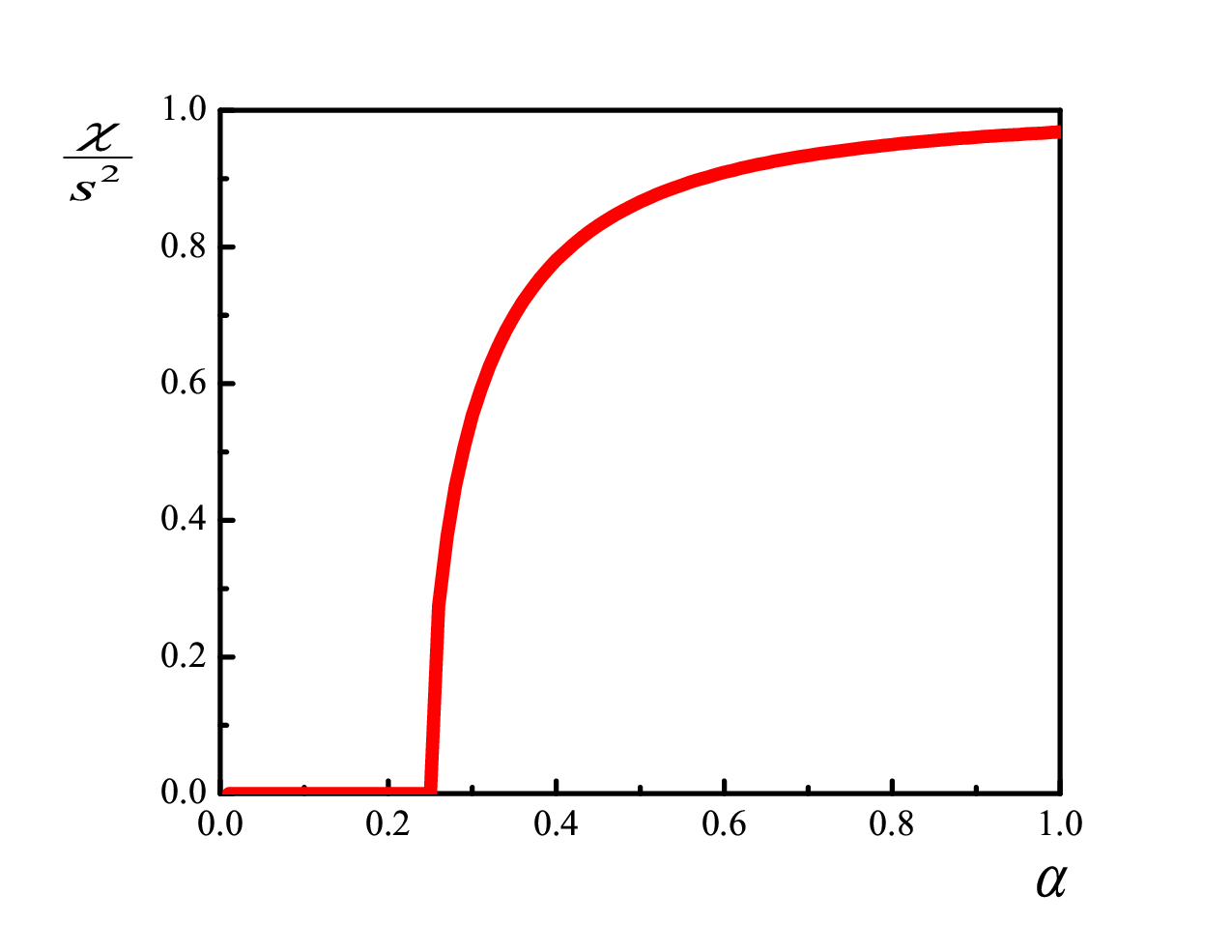}
\caption{(Color online). The  spiral order parameter as
a function of the frustration parameter $\alpha$ in the case of
non-dimerized chain ($\delta=0$).}
\label{spiral}
\end{figure}

In FIG.\ref{spiral}, we have plotted the spiral order parameter as
a function of the frustration parameter $\alpha$ for the
non-dimerized model ($\delta=0$). As is clearly seen from this
figure, there is no long-range spiral order in the region of
frustration $\alpha<\alpha_{c}=0.25$. However, in the region
$\alpha>\alpha_{c}=0.25$ the spins of the system show a profound
spiral order which grows by increasing the frustration parameter
$\alpha$.

It has been discovered that continuous phase transitions  have
many interesting properties. The phenomena associated with
continuous phase transitions are called critical phenomena, due to
their association with critical points. It turns out that
continuous phase transitions can be characterized by parameters
known as critical exponents. Critical exponents describe the
behavior of physical quantities near continuous phase transitions.
It is believed, that they are universal, i.e. they do not depend
on the details of the physical system. Our analytical results show
that the spiral order parameter $\chi$ approaches zero in a
singular fashion as $\alpha$ approaches $\alpha_{c}$, vanishing
asymptotically as

\begin{eqnarray}
\chi \propto (1-\frac{\alpha_c}{\alpha})^{1/2}, \label{exponent}
\end{eqnarray}
which shows that the critical exponent for spiral order parameter
is a simple fraction $\varepsilon=1/2$.

Now, we back to our original problem $\delta\neq0$, finding the exact ground
state phase diagram of the classical frustrated and dimerized
Heisenberg chains. One can immediately see the possibility of having two spirals,
one on the even sites, the other on the odd sites, both with
the same wave length, but with a phase difference as described in the following:

To use the cluster approach for the case of non-zero dimerization, $\delta\neq0$,
one must consider two types of cluster, one being  $(\bullet\quad \bullet\bullet)$,
the other being  $(\bullet\bullet\quad\bullet)$. Let us label the spins in the first cluster
$(S_1\quad S_2S_3)$, and the second $(S_2\quad S_3S_4)$
and continue  $(S_3\quad S_4S_5)$ , $(S_4\quad S_5S_6)$, $\cdots$.
The 2-spiral form assumes a simple spiral on the even sites, and a simple spiral
on odd sites, both with the same turn angle between spins, which we called $2\theta_0$.
Calling the angles $-\theta_0+\epsilon$ and $\theta_0+\epsilon$, respectively,  of $S_1$
and $S_3$ with the center spin  $S_2$ in the first cluster, then gives $2\theta_0$ as the angle between
$S_1$  and  $S_3$. Then, preserving the angle between $S_2$ and $S_3$ in the next cluster,
(which is now $-\theta_0-\epsilon$ , since the central spin is now $S_3$) and the angle between
$S_3$ and $S_4$ being taken as$\theta_0-\epsilon$ gives the angle $2\theta_0$ between $S_2$ and $S_4$.
Hence the first two clusters begin to show the spiral on the odd sites and the spiral
on the even sites. Continuing this to the next few clusters shows that this allows a
description of a system with two spirals, one on the odd, the other on the even sites,
both with the same turn-angle $2\theta_0$ or wavelength. Also the energy of each cluster is
the same, so one can consider just one cluster, say the first one above, in figuring
out the relation between $\epsilon$ and $\delta$. For the case of $2\theta_0=\pi/2$, spins
order as following pattern $(\nwarrow\uparrow\nearrow\longrightarrow\searrow\downarrow\swarrow\longleftarrow\nwarrow)$.
Therefor, in the case ($\delta\neq0$), the following
solutions can satisfy the minimum energy condition,
\begin{eqnarray}
(\theta,\theta')&=&(0,0),(\pi,\pi),(\pi,0),(0,\pi),\nonumber\\
(\theta,\theta')&=&(\theta_{0},-\theta_{0}+\epsilon),~~~\text{(Spiral- type)},~~~ \text{where}\nonumber\\
\cos\theta_0&=&-\frac{16\delta\alpha^2-(1-\delta)^2}{4\alpha(1-\delta)^2(1+\delta)}\nonumber.\\
\sin\theta_0&=&0\rightarrow\theta_0=n\pi\rightarrow\epsilon=0\nonumber.\\
\label{e9}
\end{eqnarray}

By substituting them into the Eq.(\ref{e4}), the ground state energy of cluster in different sectors becomes
\begin{eqnarray}
h_{\text{antiferro}}&=&h_{c}(\pi,\pi)=S^{2}(-1+\alpha),\nonumber\\
h_{\text{uud}}&=&h_{c}(0,\pi)=S^{2}(-\delta-\alpha),\nonumber\\
h_{\text{duu}}&=&h_{c}(\pi,0)=S^{2}(\delta-\alpha),\nonumber\\
h_{\text{spiral}}&=&h_{c}(\theta_{0},-\theta_{0}+\epsilon)=S^{2}\Big(\frac{\delta-1}{2}\nonumber\\
                 &+&(1+\cos\theta_0)\big\{(1-\delta)+\sqrt{(1-\delta)^2\cos^2\theta_0+4\delta}\big\}\Big).\nonumber
\label{e10}
\end{eqnarray}

As it can be seen from the above equations, in respect to the case
of $\delta=0$, dimerization exchange removed the degeneracy
between $(\downarrow,\uparrow,\uparrow)$ and
$(\uparrow,\uparrow,\downarrow)$ states. The
$(\downarrow,\uparrow,\uparrow)$ state, defines as a phase with
opposite magnetization on odd ($J(1-\delta)$) bonds, but
$(\uparrow,\uparrow,\downarrow)$ state denotes by opposite
magnetization on even ($J(1+\delta)$) bonds.
Using the conditions in Eq. (\ref{e9}) allow us to find the stability
of different phases. Doing some calculations, one can find two critical lines as

 \begin{figure}[t]
\includegraphics[width=1.1\columnwidth]{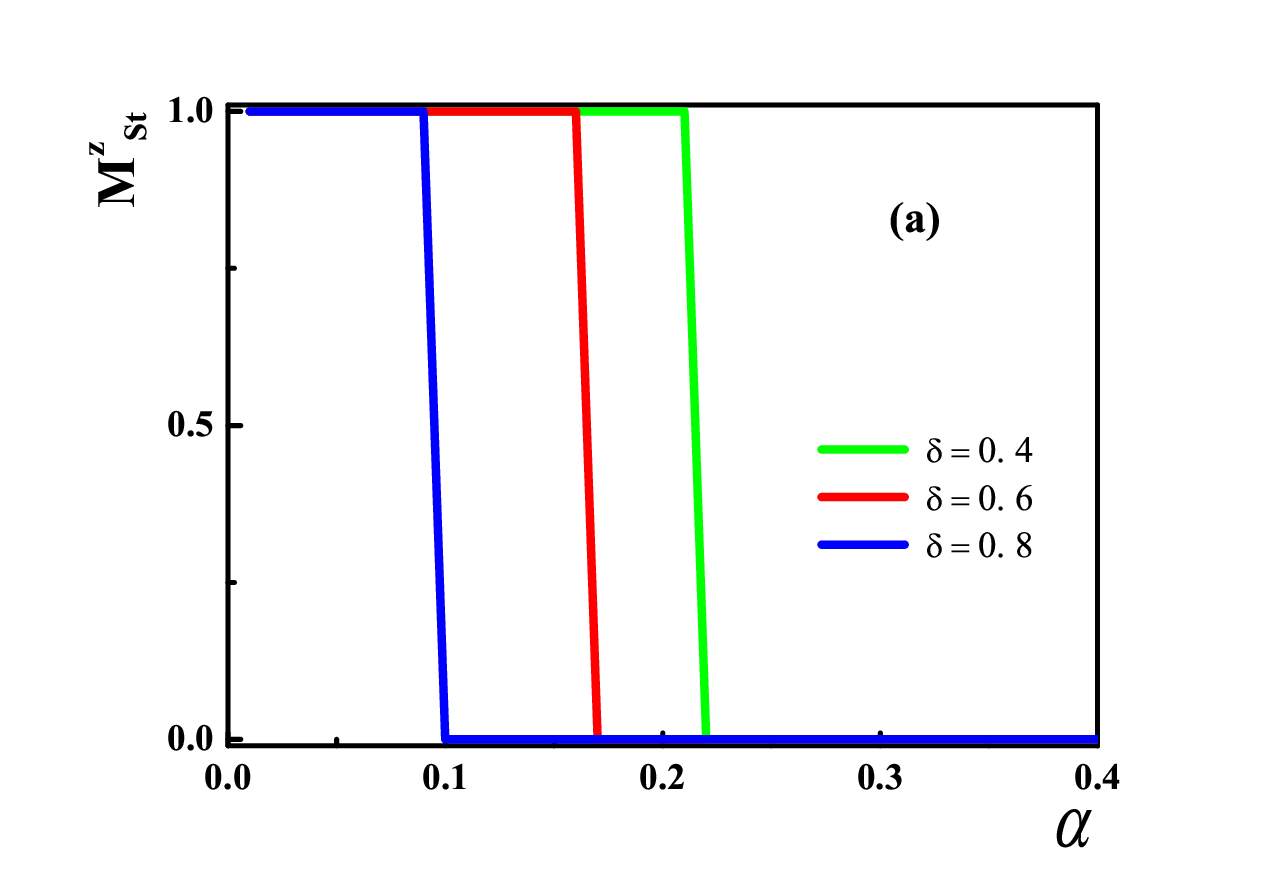}
\includegraphics[width=1.1\columnwidth]{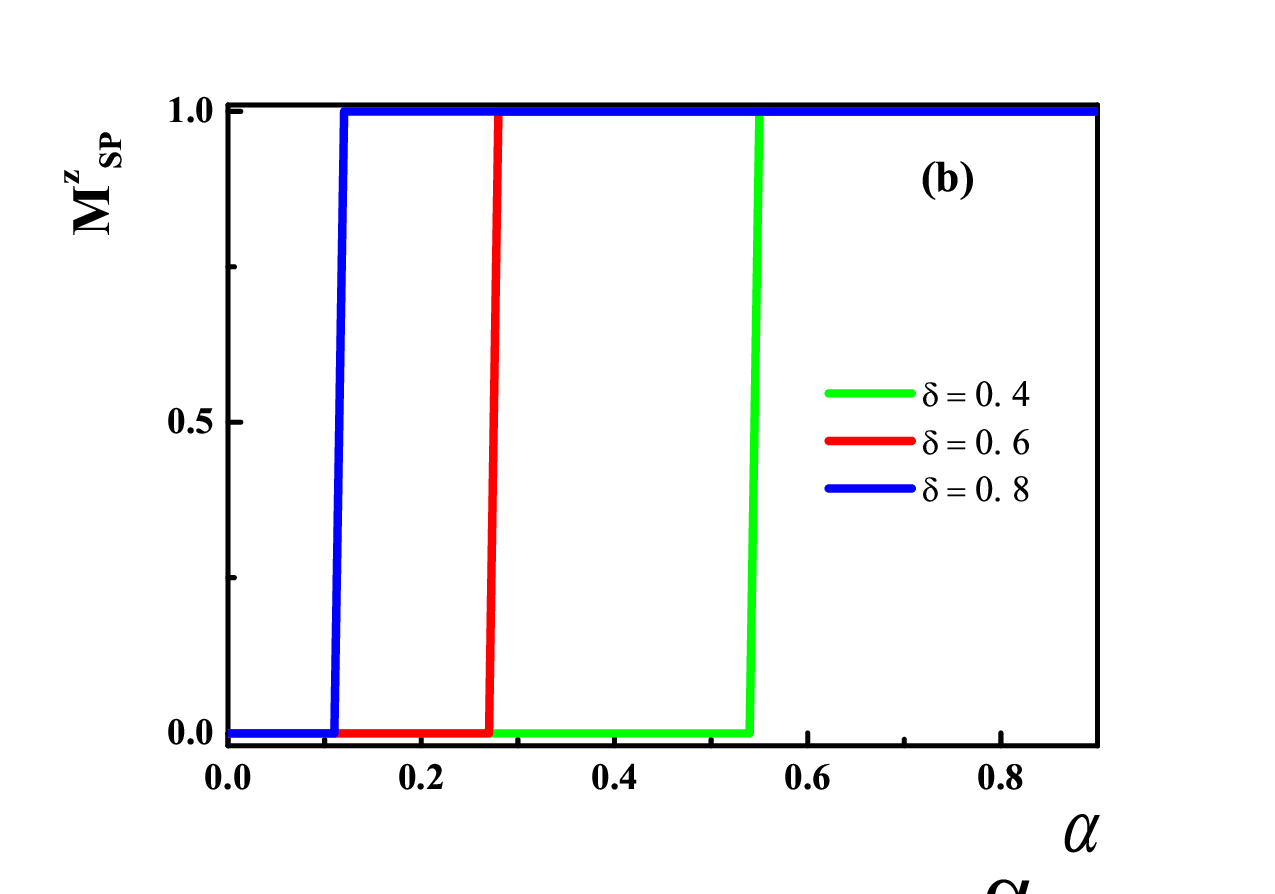}
\caption{(Color online). The staggered magnetization $M_{st}$ and stripe-antiferromagnetic
$M_{sp}$ order parameters  as function of the frustration
$\alpha$ for different values of the frustration $\delta=0.4, 0.6, 0.8$.}
\label{staggered}
\end{figure}

\begin{eqnarray}
\alpha_{\tilde{c}}&=&\frac{1-\delta^2}{4\delta}, \nonumber \\
\alpha_{\tilde{c}}&=&\frac{\delta^2-1}{4\delta}.
\label{critical-lines}
\end{eqnarray}
These critical lines separate spiral phase from the \textit{uud} and the
\textit{duu} phases. There is a discontinuity in the derivative
$\frac{\partial h_{c}}{\partial \delta}$,  and therefore a
first-order phase transition through the mentioned critical lines.
In addition, one should note that the ordering of the \textit{uud} and
\textit{duu} phases, in principle is a type of the
stripe-antiferromagnetic phase\cite{Mahdavifar08, Mahdavifar10}.
Therefore, the order parameter for distinguishing these phases is
defined as
\begin{eqnarray}
M_{sp}^{z} &=& \frac{2}{N} \sum_{n=1}(-1)^{n+1}(S_{2n-1}^{z}+
S_{2n}^{z}),~for~uud \nonumber \\
&=&\frac{2}{N} \sum_{n=1}(-1)^{n+1}(S_{2n}^{z}+
S_{2n+1}^{z}),~for~duu.
\end{eqnarray}
We have also found  antiferromagnetic phase that is stable for
\begin{eqnarray}
\alpha&<&\frac{1-\delta^2}{4}~for~\delta>0,\nonumber \\
\alpha&>&\frac{\delta^2-1}{4}~for~\delta<0.
\end{eqnarray}
It is completely clear that in \textit{uud} and \textit{duu} phases, $M_{sp}^{z}$ takes the value $1$.
In FIG.~\ref{staggered} (a), we have plotted $M_{st}^{z}$ as a function of the frustration
$\alpha$ for different values of the frustration $\delta=0.4, 0.6,
0.8$.  As it can seen from this figure, for $\alpha<\alpha_c$,
the staggered magnetization is equal to $1.0$, which shows that
the ground state of the system is in the fully polarized antiferromagnetic phase.
Finally, it is clear that there is no antiferromagnetic ordering in
the region $\alpha>\alpha_c$. In FIG.~\ref{staggered} (b), we
have the same picture for the stripe antiferromagnetic order
function. The zero value of $M_{sp}^{z}$ in the
region $\alpha<\alpha_{\tilde{c}}$ is in complete agreement with the fully
polarized antiferromagnetic and spiral phases in this region. By more increasing the
dimerization and for $\alpha>\alpha_{\tilde{c}}$, clearly be seen that
the ground state of the system is in the \textit{uud}.
Also one predicts that the tripe antiferromagnetic as a function
of the $\alpha$ displays a jump for certain parameters which is
one of the most important indications of the first-order phase transition. 
We emphasize that a first-order phase transition occurs between the spiral and the \textit{duu}
phases, for negative values of the dimerization.

\begin{figure}[t]
\includegraphics[width=1.1\columnwidth]{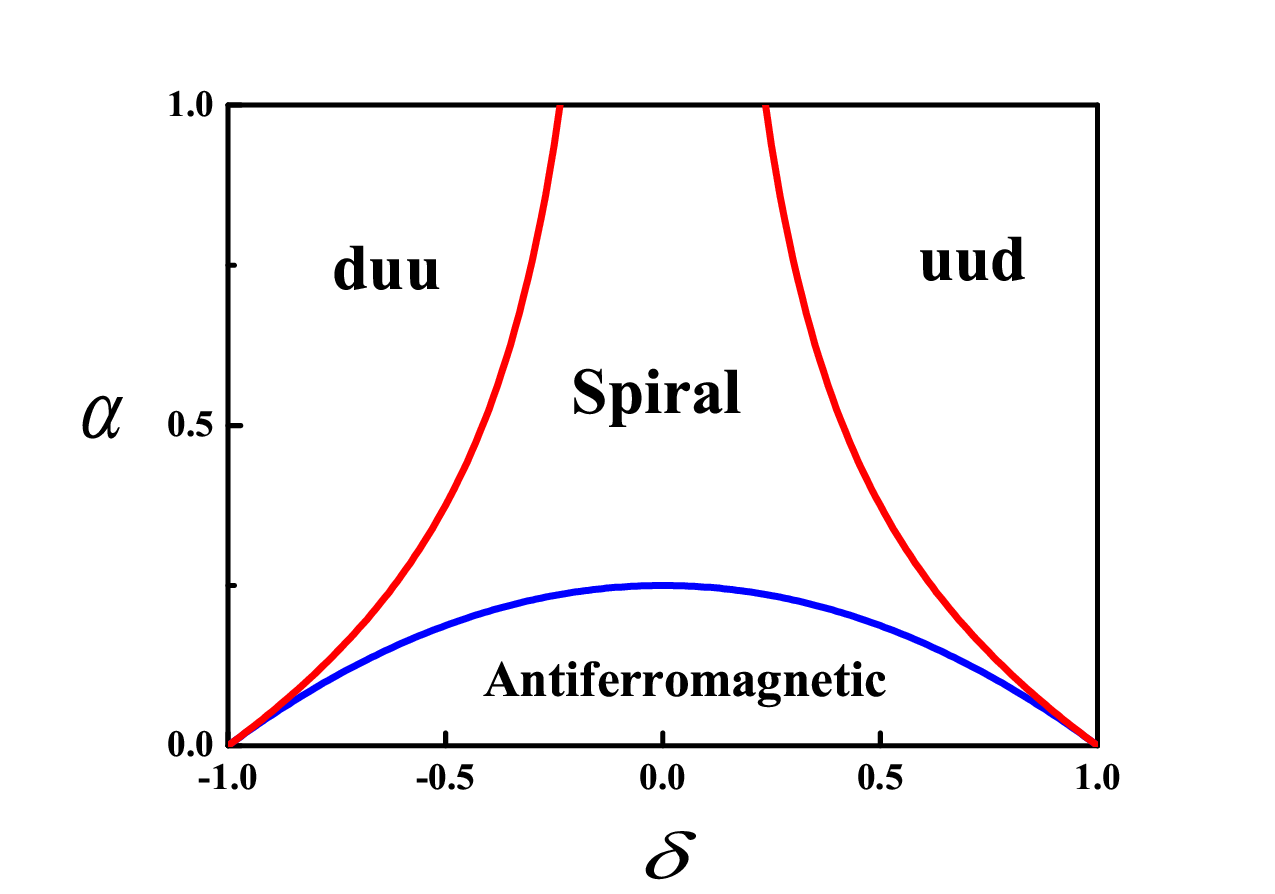}
\caption{(Color online). The classical ground state Phase diagram of the dimerized and frustrated chains in $\delta-\alpha$ plane.
The antiferromagnetic is separated from spiral  by critical line. At $\delta=0$, the spiral phase is separated from
antiferromagnetic  phass by  the critical point $\alpha=0.25$.} \label{phase}
\end{figure}

The FIG.~(\ref{phase}) shows the exact classical ground state phase
diagram of the model in $\delta-\alpha$ plane. It should be mentioned that the same
phase diagram can be also found in the ferromagnetic side which we do not depict. In the absence of
dimerization, $\delta=0$, there are antiferromagnetic(ferromagnetic) and spiral
phases which is separated by two critical point at
$\alpha_c= \pm0.25$ which the negative sign refers to ferromagnetic side. 
The second order phase transition occurs at these critical points. By
turning dimerization  the spiral phase persists to be stable in
$\frac{1-\delta^2}{4}<\alpha<\frac{1-\delta^2}{4\delta}$ region for $\delta>0$ and
$\frac{\delta^2-1}{4\delta}{4}<\alpha<\frac{1-\delta^2}{4}$ region for $\delta<0$.
The antiferromagnetic phase remains stable up to the critical line
$\frac{1-\delta^2}{4}$. The \textit{uud} and \textit{duu} phases
propagate with different energy and separated by two first order
critical lines, ($\frac{1-\delta^2}{4\delta}$ and $\frac{\delta^2-1}{4\delta}$),
from spiral phase.

In the following we are interested to implement LK method to
non-coplanar antiferromagnetic case. Again with using cluster approach and without
losing generality,  labeled the angles $(\theta,\phi)$
,$(\theta',\phi')$  made by the end spins with the central spin
(see FIG.~\ref{angel}). Then the cluster energy is determine as

\begin{eqnarray}
h_{c}(\theta,\theta')&=&S^{2}\Big\{\Big(\frac{1-\delta}{2}\Big)\cos\theta+\Big(\frac{1+\delta}{2}\Big)\cos\theta'\nonumber\\
&+&\alpha\Big[\cos\theta\cos\theta'+\sin\theta\sin\theta'\cos(\phi-\phi')\Big]\Big\}.\nonumber\\
\label{e11}
\end{eqnarray}

Minimizing over the angles $\theta,\theta',\phi$ , and $\phi'$ gives the following equations
\begin{eqnarray}
\frac{\partial h_{c}}{\partial \theta}&=&-\frac{S^{2}}{2}\Big\{(1-\delta)\sin\theta+2\alpha\big[\sin\theta\cos\theta'\nonumber\\
&-&\cos\theta\sin\theta'\cos(\phi-\phi')\big]\Big\},\nonumber\\
\frac{\partial h_{c}}{\partial \theta'}&=&-\frac{S^{2}}{2}\Big\{(1+\delta)\sin\theta'+2\alpha\big[\sin\theta'\cos\theta\nonumber\\
&-&\cos\theta'\sin\theta\cos(\phi-\phi')\big]\Big\},\nonumber\\
\frac{\partial h_{c}}{\partial \phi}&=&-\frac{S^{2}}{2}\Big\{-2\alpha\sin\theta\sin\theta'\sin(\phi-\phi')\Big\},\nonumber\\
\frac{\partial h_{c}}{\partial
\phi'}&=&-\frac{S^{2}}{2}\Big\{2\alpha\sin\theta\sin\theta'\sin(\phi-\phi')\Big\}.
\label{e12}
\end{eqnarray}

We check the possible configurations which
can minimize the above equations. By taking  arbitrary $(\phi,\phi')$, we have the solutions
$(\pi,\pi), (0,0)$ that are related to the antiferromagnetic and ferromagnetic
states respectively. We have also the solutions $(\pi,0)\rightarrow (\downarrow,\uparrow,\uparrow)$ and
$(0,\pi)\rightarrow (\uparrow,\uparrow,\downarrow)$ that are related to the \textit{uud} and \textit{duu}
states respectively. Spiral phase also exists same as the coplanar case. The stability of different phases in the
non-coplanar case is also checked and behaves same as the coplanar case.



\section{conclusion}\label{sec-III }

To summarize, we have studied  the classical ground state magnetic
phase diagram of the dimerized and frustrated Heisenberg chain
using LK cluster method. In coplanar case and in the absence of
dimerization effect this approach could detect antiferromagnetic(ferromagnetic)
and spiral phases. We have shown that turning the dimerization yields to
remove the degeneracy between two \textit{uud} and \textit{duu} phases. We have argued
that in the  ground state phase diagram of the system  there are
first order transition lines. These lines separate spiral
and \textit{uud} or \textit{duu} phases.
On the other hand two second order phase transition points also exist, which separate
antiferromagnetic(ferromagnetic) and spiral phases.
By helping this approach we have calculated the spiral exact critical exponent $\varepsilon=1/2$.

In order to generalize our treatment,
we have considered the non-coplanar case and
checked its phases by LK cluster method. Our calculations
revealed that in the non-coplanar case classical phase
diagram consist of antiferromagnetic or ferromagnetic
depend on nearest neighbor coupling, \textit{duu} and
\textit{uud} phases which are still non-degenerate
and spiral phase.
Finally, one of the main achievement of
our work which should be highlighted is that in the
both coplanar and non-coplanar cases, the spiral state is stable
in the classical phase diagram for $\delta\neq0$.




\vspace{0.3cm}


\end{document}